\documentclass [12pt]{article}
\usepackage{graphicx}
\newcommand{\cntr}[1]{{\vspace*{\fill}\vskip 2in\hspace*{\fill} #1
  \hspace*{\fill}\vspace*{\fill}}}

\begin{document}
\noindent {\bf \Large Theory of tunable pH sensitive vesicles of 
anionic and cationic lipids or anionic and neutral lipids}

\noindent Xiao-jun Li$^*$ and M. Schick 

\noindent Department of Physics, Box 351560, 
        University of Washington, Seattle, Washington 98195-1560, USA

\noindent Running title: Theory of tunable pH sensitive vesicles


\noindent \today

\begin{abstract}
The design of vesicles which become unstable at an easily tuned value of
pH is of great interest for targeted drug delivery.  We present a
microscopic theory for two forms of such vesicles.  A model of lipids
introduced by us previously is applied to a system of ionizable, anionic
lipid, and permanently charged, cationic lipid. We calculate the pH at
which the lamellar phase becomes unstable with respect to an inverted
hexagonal one, a value which depends continuously on the system
composition. Identifying this instability with that displayed by
unilamellar vesicles undergoing fusion, we obtain very good agreement
with the recent experimental data of Hafez et al., Biophys. J. 2000 79:
1438-1446, on the pH at which fusion occurs vs. vesicle composition. We
explicate the mechanism in terms of the role of the counter ions. This
understanding suggests that a system of a neutral, non lamellar forming
lipid stabilized by an anionic lipid would serve equally well for
preparing tunable, pH sensitive vesicles. Our calculations confirm
this. Further, we show that both forms of vesicle have the desirable
feature of exhibiting a regime in which the pH at instability is a
rapidly varying function of the vesicle composition.
\end{abstract}

\section{Introduction} The creation of liposomes which become unstable
in response to changes in their environment has been the object of
longstanding interest in connection with applications to drug delivery
(Yatvin et al., 1980).  One particularly interesting environmental cue
is the relatively low pH which is found in tumor tissue (Tannock and
Rotin, 1989) and in endosomes (Tycko and Maxfield, 1982). In the latter
case, the rapid acidification which occurs in the endocytic vesicle
would bring about the instability of the liposome resulting either in
the release of its contents within the endosome itself or in
liposome-mediated destabilization of the endocytic vesicle with
consequent release of the liposome's contents to the cytoplasm
(Straubinger, 1993).

There exist various strategies for producing pH-sensitive liposomes
(Thomas and Tirrell, 1992; Torchilin et al., 1993; Straubinger, 1993; 
Chu and Szoka, 1994; Sorgi and Huang, 1996).  
One method is to combine a lipid which does not
form bilayers under physiological conditions with an ionizable anionic
amphiphile or lipid.  The latter, when sufficiently charged, stabilizes
a bilayer of the combined system. The mechanism of this stabilization,
as we argue below, is the attraction of counter ions and their associated
waters of hydration to the vicinity of the headgroups which effectively
increases their size. As the pH is reduced, so is the fraction of
anionic amphiphiles which are ionized. Therefore there are fewer counter
ions near the headgroups to stabilize them. Thus the reduction in pH
eventually triggers an instability of the vesicle; the lipids revert to
their more stable phase, usually an inverted hexagonal ($H_{II}$)
one. Most commonly, the non lamellar-forming lipid is a
phosphatidyl\-ethanolamine (PE), such as dioleoyl\-phosphatidyl\-ethanolamine
(DOPE) (Cullis and de Kruijff, 1978).

One problem with this strategy is that the pH at which the instability
occurs is determined by the pK of the single ionizable component, and is
therefore not easily tuned. Discrete tuning can be obtained by utilizing
different ionizable components (Collins et al. 1989).  An alternative
method which results in a vesicle exhibiting an instability at a value
of pH which can be tuned {\em continuously} has recently been
demonstrated (Hafez et al., 2000). They utilized a vesicle with an
ionizable anionic lipid, cholesteryl hemisuccinate (CHEMS), and a
permanently charged cationic lipid N,N-dioleoyl-N,N-dimethylammonium
chloride (DODAC). The pH at which the vesicle becomes unstable is a
monotonically increasing function of the DODAC concentration, and is
therefore easily, and continuously, tuned.  We understand this result as
follows.  Vesicles consisting only of CHEMS are unstable with respect to
formation of an $H_{II}$ phase at pH less than 4.2 (Hafez and Cullis,
2000). This implies that a sufficient number of CHEMS must be ionized to
stabilize such a vesicle. These ionized headgroups attract counter ions
to their vicinity. It is these counter ions,
enlarged by their waters of hydration, that stabilize a system which
would otherwise tend to an $H_{II}$ phase.  The addition of cationic
DODAC to such a vesicle at any pH, causes a decrease in the number of
counter ions in the vicinity of the headgroups.  This decrease tends to
destabilize a previously stable vesicle. To restore the number of
counter ions near the headgroups, and the vesicle's stability, more
counter ions must be attracted to the headgroups, which can be done by
ionizing more of the CHEMS; {\em i.e.} by increasing the pH.  Hence the
value of the pH at the instability is an increasing and continuous
function of the concentration of DODAC.

In this paper, we apply our model to the system of mixed, ionizable,
anionic lipid and fully charged, cationic lipid and solvent such as that
examined by Hafez and Cullis.  We identify the pH at which bilayer
vesicles become unstable as the pH at which the lamellar phase becomes
unstable to the inverted hexagonal phase, a reasonable assumption
supported by much experimental data (Hope at al., 1983; Ellens at al.,
1986). With this identification, we indeed find that the pH at which an
instability occurs is a monotonically increasing function of the
cationic lipid concentration, or equivalently, a monotonically
decreasing function of the anionic lipid concentration. Our results fit
the experimental data very well.

The above reasoning indicates that tunable, pH sensitive liposomes
should also be formed from a mixture of a neutral lipid which favors a
non-lamellar phase, such as phosphatidyl\-ethanolamine (PE), and an
anionic lipid which stabilizes the liposome by attracting counterions to
it. Such stabilization is well known utilizing various anions, such as
palmitoylhomocysteine (Yatvin et al., 1980; Connor and Huang, 1985),
oleic acid (Straubinger et al., 1985; Wang and Huang, 1987), or CHEMS
(Straubinger, 1993; Ellens et al., 1984). As the
concentration of the stabilizing counter ions clearly depends on both
the concentration of the anionic lipid and the pH, the value of the
latter at which the vesicle becomes unstable will be tunable, depending
continuously on the concentration of anionic lipid.  To test this
hypothesis, we apply our model to a system of mixed, ionizable, anionic
lipid and neutral lipid. We again find that the pH at which an
instability occurs is a monotonically decreasing function of the anionic
lipid concentration. Our results are in accord with experiments on the
pH sensitivity of vesicles composed of
1-Palmitoyl-2-oleoyl-phosphatidyletanolamine (POPE) stabilized with
tocopherol hemisuccinate (Jizomoto et al., 1994). Lastly our results show that
both forms of vesicle have the desirable property of exhibiting a regime
in which the pH at the instability is very sensitive to the
concentration of the anionic lipid. This will always be the case
whenever a minimum amount of one lipid is required to stabilize the
formation of vesicles by the mixture.

In the following section we briefly review our model of charged lipids
(Li and Schick, 2000a) and of lipid mixtures (Li and Schick, 2000b; Li
and Schick 2000c). We
then present our results. First we show the phase diagram of a single,
ionizable, anionic lipid, solvent, and counter ions, a diagram which
shows the transition from a lamellar phase ($L_{\alpha}$) to an inverted
hexagonal one ($H_{II}$) as a function of pH. We then consider the mixed
system of ionizable anionic lipid and completely charged cationic lipid,
and show the phase behavior. There is a transition
 between $L_{\alpha}$ and $H_{II}$ phases which occurs at a value of the
pH which is a function of the concentration of the ionizable anionic
lipid. We also show the spatial distribution of the various mass and
charge densities in the coexisting phases. Lastly we consider the system
of ionizable anionic and neutral lipids, and present its phase
behavior. It also shows a transition between $L_{\alpha}$ and $H_{II}$
phases which occurs at a value of pH that depends on the concentration
of the ionizable lipid. 

\section{The Model and its Self Consistent Field Solution} 
We consider a system of volume $V$ consisting of anionic lipids, 
cationic lipids,
counter ions and solvent whose densities are controlled by the
fugacities $z_1$, $z_2$, $z_c$, and $z_s$ respectively (Li and Schick,
2000a). By taking the
charge of the cationic lipid to zero, we can also describe a mixture
of anionic and neutral lipids. 
The counter ions are positively charged. Because of
overall charge neutrality, the average amount of charge on the anionic
lipids is related to the density of cationic lipids and the density of
counterions. Hence we use the fugacity $z_c$ to control the charge on
the anionic headgroups, which is equivalent to controlling the pH.

With the exception of their Coulombic properties, the two lipids are
modelled identically. They each consist of headgroups of volume $\nu_h$,
and two equal-length, completely flexible tails each consisting of $N$
segments of volume $\nu_t$. Each lipid tail is characterized by a radius
of gyration $R_g=(Na^2/6)^{1/2}$, with $a$ the statistical segment
length. The counter ions are characterized by their charge, $+e$, and
their volume, $\nu_c$, while the neutral solvent particles 
are characterized by their
volume $\nu_s$.

There are eight local densities which specify the state of
the system. We measure them all with respect to the convenient density
$\nu_h^{-1}$. They are the number density of the headgroups of the anionic lipids,
$\nu_h^{-1}\Phi_h^{(1)}({\bf r})$, and of the cationic lipids, 
$\nu_h^{-1}\Phi_h^{(2)}({\bf r})$; 
the number density of the tail segments of each lipid,
$\nu_h^{-1}\Phi_t^{(1)}({\bf r})$ and $\nu_h^{-1}\Phi_t^{(2)}({\bf r})$;
the number density
of the solvent, $\nu_h^{-1}\Phi_s({\bf r})$ and of the counter ions 
$\nu_h^{-1}\Phi_c({\bf
r})$; and the local charge density of the headgroup of the anionic lipid,
$e\nu_h^{-1}P_h^{(1)}({\bf r})$ and of the cationic lipid, 
$e\nu_h^{-1}P_h^{(2)}({\bf r})$.
The local charge
density of the positive counter ions is simply $e\nu_h^{-1}\Phi_c({\bf r})$.
Note that all functions $\Phi({\bf r})$ and
$P({\bf r})$ are defined to be dimensionless. 

The interactions among these densities are of two kinds. First there is a
repulsive, contact interaction between headgroups and tail segments, and
also between solvent and tail segments. The strength of this interaction
is $kT\nu_h\chi$, where $k$ is Boltzmann's constant and $T$ the absolute
temperature. Second there is the Coulomb interaction between all
charges. The energy per unit volume of the system, expressed in the
natural units $kT/\nu_h$, can be written 
\begin{eqnarray}
\label{energy}
&&{\nu_h\over V kT}E[\Phi_h^{(1)}, \Phi_h^{(2)}, \Phi_t^{(1)}, 
\Phi_t^{(2)},\Phi_s, \Phi_c, P_h^{(1)}, P_h^{(2)}] \nonumber \\
&&= 2\chi N\int{d{\bf r} \over V}\left[\sum_{L=1}^2\Phi_h^{(L)}({\bf r})+
\Phi_s({\bf r})\right]
\sum_{M=1}^2\Phi_t^{(M)}({\bf r})\nonumber \\
&&+{\beta^*\over 8\pi}\int {d{\bf r}\over
V}{d{\bf r'}\over R_g^2}
\left[\sum_{L=1}^2P_h^{(L)}({\bf r})+ \Phi_c({\bf r})\right]{1\over |{\bf r}-{\bf r'}|}
\left[\sum_{M=1}^2 P_h^{(M)}({\bf r'})+\Phi_c({\bf r'})\right], 
\end{eqnarray}
where 
\begin{equation}
\label{beta}
\beta^*\equiv{4\pi e^2R_g^2\over\nu_h\epsilon kT}
\end{equation}
is a dimensionless measure of the strength of the Coulomb interaction,
and $\epsilon$ is the dielectric constant of the solvent.
In addition to these interactions, we impose a local incompressibility
constraint on the system which models the hard core interactions between
all particles. Upon defining the volume ratios $\gamma_s\equiv
\nu_s/\nu_h$, $\gamma_c=\nu_c/\nu_h$, and $\gamma_t=2N\nu_t/\nu_h$, the
incompressibility constraint that the sum of the volume fractions of all
components must be unity everywhere takes the form
\begin{equation}
\label{incomp}
\gamma_s\Phi_s({\bf r})+\gamma_c\Phi_c({\bf r})+
\sum_{L=1}^2 \left[\Phi_h^{(L)}({\bf r})+\gamma_t\Phi_t^{(L)}({\bf r})\right]=1.
\end{equation}

As shown earlier (Li and Schick, 2000a), 
the partition function of the system can
be written in the  form  
in which the eight fluctuating densities, instead of interacting
directly with one another, interact indirectly via eight fluctuating 
fields, here denoted $W_h^{(L)}$, $W_t^{(L)}$, $U_h^{(L)}$, with $L=1,2$,
and $W_s$, $U_c$. Self consistent field theory results when the fluctuating fields
and densities are approximated by those values which minimize the free energy,
$\Omega$, of the system in the presence of these fields. The free energy
to be minimized has the form
\begin{eqnarray}
\label{free1}
{\nu_h\over kTV}{\Omega} 
&=&-{1\over V}\sum_{L=1}^2z_L{\cal Q}_L[W_h^{(L)},W_t^{(L)},U_h^{(L)}]-
z_c{{\cal Q}_c[U_c]\over V}
-z_s{{\cal Q}_s[W_s]\over V} +{\nu_h\over kTV}E\nonumber \\
&-&\int{d{\bf r}\over V}[W_s({\bf r})\Phi_s({\bf r})+U_c({\bf
r})\Phi_c({\bf r}) \nonumber \\
&& \hspace*{2.5em} +\sum_{L=1}^2
\left(W_h^{(L)}({\bf r})\Phi_h^{(L)}({\bf r}) 
+W_t^{(L)}({\bf r})\Phi_t^{(L)}({\bf r})+U_h^{(L)}({\bf r})
P_h^{(L)}({\bf r})\right)]\nonumber \\
&-&\int{d{\bf r}\over V}\left[\Xi({\bf r})
\left(1-\gamma_s\Phi_s({\bf r})-\gamma_c\Phi_c({\bf r})
-\sum_{L=1}^2\Phi_h^{(L)}({\bf r})-\sum_{L=1}^2\gamma_t
\Phi_t^{(L)}({\bf r})\right)\right].
\end{eqnarray}
Here ${\cal Q}_c[U_c]$ is the
partition function of a {\em single} counter ion of unit positive charge
in an external
potential $U_c$, 
\begin{equation}
{\cal Q}_c[U_c]=\int d{\bf R}_c\exp[-U_c({\bf R}_c)],
\end{equation}
${\cal Q}_s[W_s]$ is the partition function of a
{\em single} solvent molecule in an external field $W_s$,
\begin{equation}
{\cal Q}_s[W_s]=\int d{\bf R}_s\exp[-W_s({\bf R}_s)],
\end{equation}
and
${\cal Q}_L[W_h^{(L)},W_t^{(L)},U_h^{(L)}]$, given below, is the partition
function of a {\em single} lipid of type $L$ in external fields 
$W_h^{(L)}$, $W_t^{(L)}$, and $U_h^{(L)}$. 
Note that a Lagrange multiplier $\Xi({\bf r})$ has been introduced to
enforce the incompressibility constraint of Eq. \ref{incomp}.
The functions $W_h^{(L)}$, $\Phi_h^{(L)}$ etc. which extremize this free
energy will be denoted by their corresponding lower case letters
$w_h^{(L)}$ and $\phi_h^{(L)}$ etc. 

It is not difficult to see from the form of the free energy $\Omega$ and
that of the energy $E$ of Eq. \ref{energy} that the fields acting on the
different heads and  extremizing the free energy are equal, 
$ w_h^{(1)}({\bf r})=w_h^{(2)}({\bf r})\equiv w_h({\bf r})$, that the
fields acting on the different tails and  extremizing the free energy
are equal $ w_t^{(1)}({\bf
r})=w_t^{(2)}({\bf r})\equiv w_t({\bf r})$, and that the fields acting
on all charge densities and extremizing the free energy are related, 
$u_h^{(1)}({\bf r})=u_h^{(2)}({\bf
r})=u_c({\bf r})-\gamma_c\xi({\bf r})\equiv u({\bf r}).$ Thus 
there are only five independent functions, $w_h({\bf r})$, 
$w_t({\bf r})$, $w_s({\bf r})$, $u({\bf r})$ and $\xi({\bf r})$, and
these are obtained from the five equations 
\begin{eqnarray}
\label{sc1}
 w_h({\bf r})&\equiv& 2\chi N
          \sum_{L=1}^2\phi_t^{(L)}({\bf r})+\xi({\bf r}),\\
\label{sc2}
 w_t({\bf r})&\equiv&2\chi 
          N[\phi_s({\bf r})+\sum_{L=1}^2\phi_h^{(L)}({\bf r})]+
\gamma_t\xi({\bf 
          r}),\\           
\label{sc3}
 w_s({\bf r})&=&2\chi N\sum_{L=1}^2\phi_t^{(L)}({\bf r})+\gamma_s\xi({\bf 
          r}),\\
\label{pb}
          u({\bf r})&\equiv&{\beta^*\over 
          4\pi}\int {d{\bf r}^{\prime}\over  R_g^2}{
          \sum_{L=1}^2\rho_h^{(L)}({\bf r}^{\prime})+\phi_c({\bf 
          r}^{\prime})\over|{\bf r}-{\bf 
          r}^{\prime}|},\\
\label{sc5}
          1&=&\gamma_s\phi_s({\bf r})+\gamma_c\phi_c({\bf 
          r})+\sum_{L=1}^2[\phi_h^{(L)}({\bf r})+\gamma_t\phi_t^{(L)}({\bf 
          r})].
\end{eqnarray}          
Because the field $\xi$ can be  easily eliminated, one deals essentially with 
four equations. The eight densities are all functionals of the 
above fields except $\xi$ and, therefore, close the cycle of 
self-consistent equations:
\begin{eqnarray}
\label{head}
\phi_h^{(L)}({\bf r})[w_h,w_t,u]&=
&-z_L{\delta{\cal Q}_L[w_h,w_t,u]\over\delta 
w_h({\bf r})}, \qquad {L=1,2},\\
\phi_t^{(L)}({\bf r})[w_h,w_t,u]&=&-z_L{\delta{\cal Q}_L[w_h,w_t,u]\over\delta 
w_t({\bf r})}, \qquad {L=1,2},\\
\label{headcharge}
\rho_h^{(L)}({\bf r})[w_h,w_t,u]&=&-z_L{\delta{\cal Q}_L[w_h,w_t,u]\over\delta 
u({\bf r})}, \qquad {L=1,2},\\
\label{phis}
\phi_s({\bf r})[w_s]&=&-z_s{\delta{\cal Q}_s[w_s]\over\delta w_s({\bf 
r})},\nonumber \\
                    &=&z_s\exp\{-w_s({\bf r})\},\\
\label{phic}
 \phi_c({\bf r})[u_c]&=&-z_c{\delta{\cal Q}_c[u_c]\over\delta u_c({\bf
r})},\nonumber \\  
                   &=&z_c\exp\{-u_c({\bf r})\}.
                   \end{eqnarray}
Note that one of the self-consistent equations, Eq. \ref{pb}, is  the non-linear
Poisson-Boltzmann equation, and $u({\bf r})$ is the electric potential.

With the aid of the above equations, the self consistent, or mean field, free
energy, $\Omega_{mf}$, which is the free energy function of Eq. 
\ref{free1} evaluated at the self consistent field values of the
densities and fields can be put in the form
\begin{eqnarray}
\label{omega}
-\Omega_{mf}&=&{kT\over \nu_h}\left(\sum_{L=1}^2z_L{\cal Q}_L[w_h,w_t,u]+z_c{\cal
Q}_c[u_c]+z_s{\cal Q}_s[w_s]\right) \nonumber \\
&+&
E[\phi_h^{(1)},\phi_h^{(2)},\phi_t^{(1)},\phi_t^{(2)},\phi_s,\phi_c,\rho_h^{(1)},
\rho_h^{(2)}],
\end{eqnarray}
where we have chosen $\int \xi({\bf r})d{\bf r}=0$ for convenience.
All of the above is a simple extension of the procedure in our earlier
paper (Li and Schick, 2000a) with one exception: previously we assumed the
counter ion to have neglible volume and included the interaction
between its charge and the dipole of the solvent so that it would
attract waters of hydration and gain an effective volume. Here we simply
assign the counter ion a volume, $\nu_c$, so that we need not utilize the
interaction between charges and solvent dipoles.

The fact that the anionic lipids are ionizable has the consequence that
the fields $w_h$, $w_t$ and $u$ acting on that lipid can be replaced
(Borukhov et. al. 1998; Li and Schick 2000a) by
$w_{h,eff}^{(1)}$, $w_t$, and $0$ where
\begin{equation}
w_{h,eff}^{(1)}({\bf r})=w_h({\bf r})-\ln[1+p(\exp\{u({\bf r})\}-1)].
\end{equation}
The parameter  $p$ is related to the pK of the headgroup and can therefore
be related to the fugacity $z_c$ of the counter ions by means of the 
condition of charge neutrality 
\begin{equation} 
\int d{\bf r}\left[\sum_{L=1}^2\rho_h^{(L)}({\bf r})+\phi_c({\bf r})\right]=0.
\end{equation}
In practice, we
use this fugacity to control the fractional charge on the 
anionic lipid and therefore the pH.

In the system in which the cationic lipid is fully charged, 
the fields $w_h$, $w_t$ and $u$ acting on it can be replaced by 
$w_{h,eff}^{(2)}$, $w_t$, $0$ with
\begin{equation}
w_{h,eff}^{(2)}({\bf r})=w_h({\bf r})+u({\bf r}).
\end{equation}
Note that from Eqs. \ref{head} and \ref{headcharge} it immediately
follows that the number density of the headgroup and its charge density,
in units of $e$, are identical,  
\begin{equation}
\label{rhoh2}
\rho_h^{(2)}({\bf r})=\phi_h^{(2)}({\bf r}),
\end{equation}
as they should be because the cationic lipid is always fully charged.

In the other system which we consider, the second lipid is neutral so that
\begin{equation}
w_{h,eff}^{(2)}({\bf r})=w_h({\bf r}),
\end{equation}
and
\begin{equation}
\label{rho22}
\rho_h^{(2)}({\bf r})=0.
\end{equation}
 
There remains only to specify how
the partition function of the lipids
is calculated. As in our earlier study (Li and Schick, 2000a),
one defines the end-segment distribution function 
$q^{(L)}({\bf r},s)$ which satisfies the equation
\begin{equation}
{\partial q^{(L)}({\bf r},s)\over \partial s}=2R_g^2\nabla^2q^{(L)}({\bf
r},s)
-\left[w_{h,eff}^{(L)}({\bf r})\delta(s-1/2)+w_t({\bf
r})\right]q^{(L)}({\bf r},s),
\end{equation}
with initial condition
\begin{equation} 
q^{(L)}({\bf r},0)=1.
\end{equation}
From this function, one obtains the partition functions of the lipids, 
\begin{equation}
{\cal Q}_L=\int d{\bf r}\ q^{(L)}({\bf r},1), 
\end{equation} 
the head and tail
densities
\begin{eqnarray}
\label{phih}
\phi_h^{(L)}({\bf r})&=&\exp\{-w^{(L)}_{h,eff}({\bf r})\}q^{(L)}\left({\bf r},{1\over
2}-\right)
q^{(L)}\left({\bf r},{1\over 2}-\right),\qquad L=1,2,\\
\label{phit}
\phi_t^{(L)}({\bf r})&=&\int_0^1ds\ q^{(L)}({\bf r},s)q^{(L)}({\bf
r},1-s),\qquad L=1,2,
\end{eqnarray}
and the charge density of the anionic lipid head
\begin{equation}
\label{rhoh1}
\rho_h^{(1)}({\bf r})=-{p(z_c)\exp\{u({\bf r})\}\over1+p(z_c)(\exp\{u({\bf
r})\}-1)}\phi_h^{(1)}({\bf r}).
\end{equation}
The average fractional charge, $f_c$,
on the anionic lipid headgroup follows  
\begin{equation}
f_c(z_c)\equiv-{\int \rho_h^{(1)}({\bf r}) d{\bf r}\over 
\int \phi_h^{(1)}({\bf r}) d{\bf r}},
\end{equation}
from which  
the pH relative to the pK of the anionic lipid headgroup is obtained:
\begin{equation}
pH=pK +\log_{10}\left({f_c\over 1-f_c}\right).
\end{equation}
To summarize: there are five self-consistent equations to be solved for
the five fields $w_h({\bf r})$, $w_t({\bf r})$, $w_s({\bf r})$, $u({\bf
r})$, and $\xi({\bf r})$. They are Eqs. \ref{sc1} to \ref{sc5}. 
The fields depend on the eight densities $\phi_h^{(L)}({\bf r})$,
$\phi_t^{(L)}({\bf r})$, $\rho_h^{(L)}({\bf r})$ with $L=1,2$, $\phi_s({\bf r})$, and 
$\phi_c({\bf r})$ which depend, in turn, on these fields. The densities 
are given by Eqs.  
\ref{phis}, \ref{phic}, \ref{rhoh2} or \ref{rho22}, \ref{phih}, \ref{phit}, and \ref{rhoh1}. 
Once the fields and densities are obtained, the free energy follows 
from Eq. \ref{omega}.

Instead of solving these equations in real space, we do so in Fourier
space in such a way as to guarantee that our solution has the symmetry of
either the lamellar or inverted hexagonal phases (Matsen and Schick,
1994). Comparison of the free
energies of these phases tells us which is the globally stable one. We
do this for different temperatures, lipid concentrations and pH, and
thereby map out the phase diagram.
\section{Results}

We first consider the system of the single anionic lipid in solvent. The
architecture of this lipid is characterized by  $\gamma_t$, the
ratio of the volume of its tail groups to the volume of its head.  In
choosing the value of this parameter, we have been guided first and
foremost by the requirement that our model lipid exist in the inverted
hexagonal phase when its head group is neutral and the system is
hydrated in order that it model correctly the behavior of CHEMS (Hafez
and Cullis, 2000).  The value we have chosen, $\gamma_t=2.5$ does indeed
produce a model lipid which exists in the inverted hexagonal phase over
a large region of phase space when it is neutralized, as is seen below.
We note in passing that this is not an unreasonable value when compared
to that which follows from volumetric data on the non-lamellar forming
lipid DOPE (Rand and Fuller, 1994), $\gamma_t=2.94$, a value which most
likely assigns some of the volume of the waters of hydration, those most
tightly bound, to the headgroup.  But in our model, some of the volume
of water of hydration should be included in the volume of the head group
because the only interaction it has with water occurs when it has a net
charge.  The model interactions, therefore, neglect those waters
attracted via their dipole moment to the charges of a physical, neutral,
head group.
 
The solvent is characterized by its relative volume
$\gamma_s\equiv \nu_s/\nu_h=0.1$, close to the ratio of 0.096 appropriate
to water and a phosphatidylethanolamine headgroup (Rand and Fuller,
1994; Kozlov et al. 1994). The counter ions are modeled as H$_9$O$_4^+$,
a reasonable choice (Bell, 1959), and are therefore
characterized by their relative volume $\gamma_c\equiv \nu_c/\nu_h=0.4.$ 
The strength of the Coulomb interaction is, again, given by the
parameter $\beta^*$, Eq. \ref{beta}, which can be written as
$\beta^*=\xi/L_1$, where $\xi\equiv e^2/\epsilon kT$ is the Bjerrum
length, and $L_1\equiv \nu_h/4\pi R_g^2$ is a length characterizing the
architecture of lipid 1. Using the value of $R_g$ found earlier (Li and
Schick, 2000a) to be appropriate to DOPE and a  Bjerrum length of 7
\AA\  appropriate for water, we obtain $\beta^*=27$ and have used this value.
The phase diagram of the system is shown in
Fig. 1 as a function of the effective temperature $T^*\equiv (2\chi
N)^{-1}$ and the pH relative to the pK of the lipid headgroup.
We observe the characteristic transition between the inverted hexagonal
and lamellar phases as the pH is increased (Hope and Cullis, 1980;
Bezrukov et al., 1998).  The fugacity of solvent
here is $z_s=3.2$. At this value, the system is not in the presence of
excess water.  When completely neutralized, the two phases coexist at
$T^*=0.065$ with the $H_{II}$ phase containing a volume fraction of
solvent $\gamma_s\phi_s=0.063$ and the lamellar phase a volume fraction
$\gamma_s\phi_s=0.087$. For a given temperature, the transition between
phases occurs at a given pH. If this value is not a biologically useful
one, as it is not for PS which undergoes a phase transition at the very
acidic value pH $\approx$ 3 (Hope and Cullis, 1980), then a vesicle
made from this lipid is not applicable for drug delivery.

To vary continuously 
the value of the pH at which the lamellar phase becomes unstable for a fixed
temperature, one can add an additional lipid. We now consider the case
when this additional lipid is cationic, and fully charged as in the
experiment of Hafez et al. (Hafez et al., 2000). The phase diagram we
obtain for this system is shown in Fig. 2 as a function of the
fractional concentration $\Theta$ of the ionizable anionic lipid 
\begin{equation}
\Theta={\phi_h^{(1)}+\gamma_t\phi_t^{(1)}\over
\sum_{L=1}^2[\phi_h^{(L)}+\gamma_t\phi_t^{(L)}]},
\end{equation}
{\em vs.} pH $-$ pK, where the pK is that of the anionic lipid headgroup. Our
results for the phase coexistence are shown in solid lines. We have
chosen the solvent activity $z_s=3.425$ and the effective temperature to
be $T^*=0.079$. Under these conditions, the system of pure anionic lipid
is, when completely neutralized, just at
phase coexistence between $L_{\alpha}$ and $H_{II}$ phases. 
Therefore  our curves, in the limit of very large
negative pH, asymptote to $\Theta=1$. This coexistence is characterized
by a volume fraction of solvent $\gamma_s\phi_s=0.087$ in
the inverted hexagonal phase and $\gamma_s\phi_s=0.116$ in the lamellar
phase. Because the phase boundary curves are flat near $\Theta=1.0$, 
our results are not very sensitive to different choices of temperature
and solvent chemical potential provided, of course, that we remain in
the same general region of phase behavior. The experimental results   
of Hafez et al., which show the pH at which vesicle fusion occurs (their
pH$_f$), are shown by the solid dots. To obtain the best fit, we have 
taken the pK of CHEMS to be
5.5. The actual value of the pK of CHEMS in the DODAC/CHEMS system has
not been measured. It has been measured in a large unilamellar vesicle
composed of CHEMS and DOPE and a value of 5.8 obtained (Hafez and
Cullis, 2000). Thus the value of 5.5 we have used in our fit to the data
is not unreasonable. We note that our results fit the data rather well
except for the very largest value of pH. Given the assumptions in the
modelling, and the assumption that the 
architecture of the two lipids is the same, this agreement is
gratifying. We also restate the interesting point made by Hafez et
al. that the preferred phase of the lipid mixture can be inverted
hexagonal even though both of the lipids in isolation adopt a 
lamellar organization. They do so, in our view, because in isolation
their headgroups are sufficiently charged to attract stabilizing counter
ions which increase the effective size of their headgroups.
The transition to the inverted hexagonal phase comes about because, as
the lipids are mixed, the number of those stabilizing counter ions is
reduced until eventually the lamellar phase becomes unstable.

A somewhat different point of view is taken by Hafez et al. They assume
that for fusion to occur, {\em i.e} the instability, the surface charge
on the vesicle must be zero, permitting close contact. Therefore the
proportion of CHEMS which is negatively charged must equal the DODAC
content of the membrane. Equivalently, the number of counter ions must
be identically zero. This condition is expressed in the equation
\begin{equation}
\label{ismail}
pH-pK=\log_{10}\left({1-\Theta\over 2\Theta-1}\right), 
\end{equation}
which is plotted in Fig. 2 as the dashed line.

The spatial distributions of the various components and of the charges
is shown in Fig. 3 for the lamellar phase and Fig. 4 for the inverted hexagonal
phase which coexist near pH$-$pK=0. In part (a) of each figure, the volume
fractions of all elements are shown: of the headgroup of the anionic
lipid, $\phi_h^{(1)}$, of the tails of the anionic lipid,
$\gamma_t\phi_t^{(1)}$, of the headgroup of the cationic lipid,
$\phi_h^{(2)}$, and of the tails of the cationic lipid,
$\gamma_t\phi_t^{(2)}$, of the solvent, $\gamma_s\phi_s$, and of the
counter ions, $\gamma_c\phi_c$. In part (b) of each figure, the charge
distributions, in units of $e\nu_h^{-1}$, are shown: that of the anionic
lipid, $\rho_h^{(1)}$, of the cationic lipid, $\rho_h^{(2)}$, of the counter
ions, $\phi_c$, and the total charge distribution. 
It should be recalled that these distributions are those of a
lamellar phase, not an isolated lipid bilayer. 
In the latter,
the volume fraction of headgroups would fall rapidly in the 
solvent-rich regions on either side of the bilayer.
The points $x=0$ and $x=D$ correspond to the centers of sequential solvent
regions within the lamellae, with $D=2.94R_g$ being the lamellar
period. 
In Fig. 4, $x=0$
and $x=D$ correspond to the centers of adjacent tubes with $D=3.03R_g$
the lattice constant of the inverted hexagonal phase. As noted
previously (Li and Schick, 2000a), a single dielectric constant has, for
simplicity, been employed for the entire system. Were a different
dielectric constant employed in the tail region, the distribution of
counter ions would change, with less of them in the tail region. However
their volume fraction in that region is already small due to their
non-zero volume and the incompressibility constraint. Hence any change
in the distribution would probably be small.

We turn now to the system of anionic and neutral lipids. The phase
diagram is shown in Fig. 5. As in Fig. 2, $\Theta$ is the fractional
composition of the anionic lipid. The temperature $T^*$ and solvent
fugacity are the same as in Fig. 2. The results for the phase
coexistence of the anionic, neutral
lipid system are shown with solid lines. They are compared to the
calculated results
for the anionic, cationic system shown previously in Fig. 2, and 
repeated here in dashed-dotted lines.
We see again in this system of anionic and neutral lipid the
characteristic transition from $H_{II}$ to $L_{\alpha}$ phases with
increasing pH.  Again the pH at which the transition occurs is a
continuous function of the system's composition, decreasing with an
increase in the composition of the ionizable anionic lipid. This is in
agreement with results on vesicles of POPE and CHEMS (Jizomoto et al., 1994).
Equally important, we note a region in which a very
small change in the composition of the system brings about a large
change in the pH at which the instability of the lamellar phase
occurs. We also note that a minimum fractional composition
of anionic lipid, approximately 0.3, is necessary
to stabilize the DOPE-like neutral lipid. This is similar to the
experimental observation that a minimum molar composition of 0.2 CHEMS 
was necessary to stabilize vesicles of transesterified egg
phosphatidylethanolamine (Lai et al., 1985). It is easy to see from
Fig. 5 that the existence of a minimum concentration of stabilizing
anionic lipid guarntees a regime in which the pH at the instability
changes rapidly with concentration.
We understand the instability in this system 
in the same manner as for the previous
system. The neutral lipid prefers to be in an inverted hexagonal
phase. The addition of an anionic lipid can stabilize a lamellar phase
of the mixture if
there is enough of it, and if these lipids are sufficiently charged. It
does so by attracting a sufficient number of counter ions which
effectively increase the headgroup of the anionic lipid. When the pH is
changed so that fewer counter ions are attracted, the lamellar phase is
less stable, and eventually becomes unstable with respect to the
inverted hexagonal phase. Note that the criterion of Hafez et al. that
the density of counter ions be zero at the transition 
yields no useful information for this system. 

\section{Discussion}

We have applied our model of lipids, introduced previously (Li and
Schick, 2000a), to the systems of anionic lipids mixed either with a
cationic lipid or a neutral lipid. We have solved the model within self
consistent field theory, and obtained the phase diagram for these
systems showing the transition from the lamellar to inverted hexagonal
phases which occurs at a value of the pH which depends continuously 
on the membrane composition. Identifying the instability of unilamellar
vesicles as evidenced by their fusion with the instability of the
lamellar phase of the lipid mixture, we obtain good agreement between
our calculations and the experimental data of Hafez et al. for the mixed
anionic, cationic system which they studied (Hafez et al., 2000). We
have interpreted the instability in terms of the counter ions which, by
increasing the effective size of the anionic lipid, stabilize the
lamellar phase of the mixture. As cationic lipids are added, 
counter ions are subtracted. When their number gets too low, the vesicle
becomes unstable, and the lipids try to revert to an inverted hexagonal
phase. As noted by Hafez et al., the pH at which this instability occurs is a continuous function
of the composition of the vesicle, and can therefore be readily tuned to
occur at a biologically relevant value of pH. Hence such a vesicle has
the possibility of being used for drug delivery. Applying similar
reasoning, we suggested that a vesicle consisting of a neutral,
non-lamellar forming  lipid, like PE, and stabilized by the presence of 
an anionic lipid could also be used to make pH sensitive
vesicles whose pH at the point of instability could also be readily tuned.
We applied our model to such a system and found this to be true, and
that both sorts of vesicles displayed a region in which the pH at
instability was a sensitive function of composition, and thus readily
tuned. 

It might be argued that the instability of the mixed, charged, lipid vesicles 
prepared by Hafez et al. does not reflect the $L_{\alpha}$ to $H_{II}$ 
phase transition of the  bulk system, but rather is induced by a phase 
separation of the components. Such a scenario would be contrary to 
experimental evidence previously cited (Hope et al., 1983; Ellens et al., 
1986) and seems to us most unlikely. Not only is the mixed system 
energetically favorable, due to the Coulomb attraction of the two components, 
but it is also entropically favorable because formation of the mixture 
liberates the counter ions needed to neutralize the phase-separated 
charged components (Paulsen et al., 1988).

Throughout this paper we have reiterated the view that the counter ions,
which we have treated as a proton with four waters of hydration,
H$_9$O$_4^+$, play an important role in bringing about the transition
with pH by effectively increasing the volume of the headgroup of the
anionic lipid. This is reasonable given the well-known stabilization of
the lamellar phase with respect to the inverted hexagonal one as the
headgroup volume increases (Gruner, 1989). We further understand the
importance of the counter ions as follows. The Coulomb interaction in
this system has several effects. One of them is that just noted;
the charged headgroups attract counter ions which have gained an
effective volume due to their interaction with the dipoles of water, and
these tend to  stabilize the lamellar phase with respect to the inverted
hexagonal.  However the Coulomb interaction has another effect, which is
less appreciated. The Coulomb repulsion between the headgroups tends to
separate the lipids, just as an increase in temperature does, and the
consequence is the same; the tails have more room in which to move, and
this tends to {\em destabilize} the lamellar phase with respect to the inverted
hexagonal one. Thus there is a competition between these two effects,
the outcome of which depends, among other things, on the effective
volume of the stabilizing counter ions.  To confirm this picture, we
have verified by explicit calculation that, if the effective volume of
the counter ions
is too small, the destabilizing tendency of the
repulsion between headgroups overcomes the stabilizing tendency of the
added volume of counter ions, with the result that the system makes a
transition from lamellar to inverted hexagonal on increasing the
pH. This is, of course, opposite to experimental observation and to our
calculations using counter ions of volume appropriate to those found in
water.
Nonetheless, the existence of this competition between different aspects
of the Coulomb interaction 
emphasizes the importance of the counter
ions and their volume, and indicates at least one mechanism whereby 
the pH at which the lamellar
phase becomes unstable should be expected to depend not only upon the system
composition, but also upon the species of counter ion and the nature of
the solvent, as is observed (Seddon, 1990).

We are grateful to Dr. I.M. Hafez for providing a preprint of his work
with S. Ansell and P.R. Cullis, and for informative correspondence. 
This work was supported in part by the National Science Foundation under
grant number DMR9876864.

\newpage
\begin{center}
{\bf REFERENCES}
\end{center}
\begin{description}
\item{*} Current address: The Institute for Systems Biology, 
	4225 Roosevelt Way NE, Suite 200, Seattle, Washington 98105-6099, USA. 
\item{} Bell, R.P. 1959. The Proton in Chemistry. Cornell University
Press, Ithaca.
\item{} Bezrukov, S.M., R.P. Rand, I. Vodyanoy, and
V.A. Parsegian. 1998. Lipid packing stress and polypeptide aggregation:
alamethicin channel probed by proton titration of lipid charge. {\em
Faraday Discuss.} 111:173-183.
\item{} Borukhov, I., D. Andelman, and H. Orland. 1998. Random
  polyelectrolytes and polyampholytes in solution. {\em Eur. Phys. J. B} 
5:869-880.
\item{} Chu, C.-J., and F.C. Skoza. 1994. pH-sensitive liposomes. {\em
J. Liposome Res.} 4:361-395.
\item{} Collins, D. F. Maxfield, and L. Huang. 1989. Immunoliposomes with
different acid sensitivities as probes for the cellular endocytic
pathway. {\em Biochim. Biophys. Acta.} 987:47-55.
\item{} Connor, J. and L. Huang. 1985. Efficient cytoplasmic delivery of
a fluorescent dye by pH-sensitive immunoliposomes. {\em J. Cell Biol.}
101:582-589.
\item{} Cullis, P.R. and B. de Kruijff. 1978. The polymorphic phase
behavior of phosphatidyl\-ethanolamines of natural and synthetic origin. A
$^{31}$P NMR study. {\em Biochim. Biophys. Acta.} 513:31-42.
\item{} Ellens, H., Bentz, J. and F.C. Skoza. 1984. pH-induced
destabilization of phosphatidyl\-ethanolamine-containing liposomes: role
of bilayer contact. {\em Biochemistry} 23:1532-1538.
\item{} Ellens, H., Bentz, J. and F.C. Skoza. 1986. Destabilization of
phosphatidyl\-ethanolamine liposomes at the hexagonal phase transition
temperature. {\em Biochemistry} 25:285-294.
\item{} Gruner, S.M. 1989. Stability of lyotropic phases with curved
interfaces. {\em J. Phys. Chem.} 93:7562-7570. 
\item{} Hafez, I.M. and P.R. Cullis. 2000. Cholesteryl hemisuccinate
exhibits pH sensitive polymorphic phase behavior. {\em
Biochim. Biophys. Acta} 1463:107-114.
\item{} Hafez, I.M., S. Ansell, and P.R. Cullis. 2000. Tunable pH
sensitive liposomes composed of mixtures of cationic and anionic
lipids. {\em Biophys. J.} 79:1438-1446.
\item{} Hope, M.J. and P.R. Cullis. 1980. Effects of divalent cations
and pH on phosphatidylserine model membranes: A $^{31}$P NMR study {\em
Biochem. Biophys. Res. Comm.} 92:846-852.
\item{} Hope, M.J., D.C. Walker, and P.R. Cullis. 1983. Ca$^{2+}$ and pH
induced fusion of small unilamellar vesicles consisting of
phosphatidyl\-ethanolamine and negatively charged phopholipids: a freeze
fracture study. {\em Biochem. Biophys. Res. Commun} 110:15-22.
\item{} Jizomoto, H., E. Kanaoka, and K. Hirano. 1994. pH-sensitive
liposomes composed of tocopherol hemisuccinate and of
phosphatidyl\-ethanolamine including tocopherol hemisuccinate. {\em
Biochim. Biophys. Acta} 1213:343-348.
\item{} Kozlov, M.M., S. Leikin, and R.P. Rand. 1994. Bending, hydration,
and interstitial energies quantitatively account for the
hexagonal-lamellar-hexagonal re-entrant phase transition in
dioleoylphosphatidyl\-ethanolamine. {\em Biophys. J.} 67:1603-1611.
\item{} Lai, M.-Z., W.J. Vail, and F.C. Szoka. 1985. Acid- and
calcium-induced structural changes in phosphatidyl\-ethanolamine membranes
stabilized by cholesteryl hemisuccinate. {\em Biochemistry} 24:1654-1661.
\item{} Li, X.-J., and M. Schick. 2000a. Theory of lipid polymorphism:
application to phosphatidyl\-ethanolamine and
phosphatidylserine. {\em Biophys J.} 78:34-46.
\item{} Li, X.-J., and M. Schick. 2000b. Distribution of lipids in
nonlamellar phases of their mixtures. {\em J. Chem. Phys.}
112:6063-6072.
\item{} Li, X.-J., and M. Schick. 2000c. Fluctuations in mixtures of
lameller- and nonlameller-forming lipids. {\em J. Chem. Phys.}
112:10599-10607.
\item{} Matsen, M.W. and M. Schick. 1994. Stable and unstable phases of a
diblock copolymer melt. {\em Phys. Rev. Lett.} 72:2660-2663.
\item{} Paulsen, M.D., C.F. Anderson, and M.T. Record. 1988. Counterion
exchange reactions on DNA: Monte Carlo and Poisson-Boltzmann
analysis. {\em Biopolymers} 27:1249-1265.
\item{} Rand, R.P. and N.L. Fuller. 1994. Structural dimensions and
their changes in a re-entrant hexagonal-lamellar transition of
phospholipids. {\em Biophys. J.} 66:2127-2138.
\item{} Seddon, J.M. 1990. Structure of the inverted hexagonal phase
(H$_{II}$) phase, and non-lamellar phase transitions of lipids. {\em
Biochim. et Biophys. Acta.} 1031:1-69.  
\item{} Sorgi, F.L. and L. Huang. 1996. Large scale production of
DC-Chol liposomes by microfluidization. {\em Int. J. Pharm.} 144:131-139.
\item{} Straubinger, R.M. 1993. pH-sensitive liposomes for delivery of
macromolecules into cytoplasm of cultured cells. {\em Methods Enzymol.}
221:361-376.
\item{} Straubinger, R.M., Duzgunes, N., and D. Papahadjopoulos. 1985. 
pH-sensitive liposomes mediate cytoplasmic delivery of encapsulated
macromolecules. {\em FEBS Lett.} 179:148-154.
\item{} Tannock, I.F. and Rotin, D. 1989. Acid pH in tumors and its
potential for therapeutic expolitation. {\em Cancer Res.} 49:4373-4384.
\item{} Thomas, J.L. and D.A. Tirrell. 1992. Polyelectrolyte-sensitized
phospholipid vesicles. {\em Acc. Chem. Res.} 25:336-342.
\item{} Torchilin, V.P., F. Zhou, and L. Huang. 1993. pH-sensitive
liposomes. {\em J. Liposome Res.} 3:201-255.
\item{} Tycko, B. and Maxfield F.R. 1982. Rapid acidification of endocytic
vesicles containing $\alpha_2$-Macroglobulin. {\em Cell} 28:643-651.
\item{} Wang, C.Y., and L. Huang. 1987. pH-sensitive immunoliposomes
mediate target-cell-specific delivery and controlled expression of a
foreign gene in mouse. {\em Proc. Natl. Acad. Sci. USA} 84:7851-7855.  
\item{} Yatvin, M.B., Kreutz, W., Horwitz, B.A., and Shinitzky,
M. 1980. pH-sensitive liposomes: possible clinical implications. {\em
Science} 210:1253-1255.

\end{description}

\newpage
\begin{center}
{FIGURE LEGENDS}
\end{center}
\begin{description}
\item[Figure 1] Phase diagram in the temperature, $T^*$, pH plane for a
system of a single anionic lipid, solvent, and counter ions. 
The pK is that of the
anionic headgroup. The volume of the headgroup relative to that of the
entire lipid is 0.286, similar to that of DOPE, the relative volume of the
solvent is close to that of water, and the relative volume of the
counter ions is that of $H_9O_4^+$.

\item[Figure 2] Phase diagram, at fixed temperature, $T^*=0.079$  and solvent
activity, $z_s=3.425$, of a mixture of ionizable, anionic lipid,
and fully ionized cationic lipid. The volume  of the headgroup of each
lipid relative to that of the entire lipid is the same as that in
Fig. 1. The relative concentration of the ionizable lipid is denoted
$\Theta$. The solid lines show the coexistence obtained from our
calculation, the solid circles show the data from Hafez et al., 2000,
and the dashed line their criterion of vanishing counter ion density,
Eq. \ref{ismail}.  
\item[Figure 3] For the system of Fig. 2, the spatial distributions of the various components 
and of the charges
is shown for the lamellar phase at coexistence with the inverted
hexagonal phase near pH$-$pK=0. Part (a); shown are the volume
fractions of the headgroup of the anionic
lipid, $\phi_h^{(1)}$, of the tails of the anionic lipid,
$\gamma_t\phi_t^{(1)}$, of the headgroup of the cationic lipid,
$\phi_h^{(2)}$, and of the tails of the cationic lipid,
$\gamma_t\phi_t^{(2)}$, of the solvent, $\gamma_s\phi_s$, and of the
counter ions, $\gamma_c\phi_c$. Part (b); shown are the charge
distributions, in units of $e\nu_h^{-1}$, of the anionic
lipid, $\rho_h^{(1)}$, of the cationic lipid, $\rho_h^{(2)}$, of the counter
ions, $\phi_c$, and the total charge distribution. The points $x=0$ and
$x=D$ correspond to the centers of adjacent solvent regions, 
with $D=2.94R_g$ being the
lamellar period. 
\item[Figure 4] The same quantities which are shown in Fig. 3a and 3b for the
lamellar phase are shown here for the inverted hexagonal phase with
which the lamellar phase coexists. The points $x=0$ and $x=D$ correspond
to the centers of adjacent tubes, with $D=3.03 R_g$.
\item[Figure 5] Phase diagram, at fixed temperature, $T^*=0.079$ and
solvent activity, $z_s=3.425$, of a mixture of ionizable, anionic lipid,
and neutral lipid. The volume of the headgroup of each lipid relative to
that of the entire lipid is the same as that in Fig. 1. The relative
concentration of the ionizable lipid is denoted $\Theta$. The solid
lines show the coexistence obtained from our calculation. For
comparison, the dashed dotted lines show the calculated coexistence for
the ionizable anionic and fully ionized cation lipid system shown
previously in Fig. 2 as solid lines.

\end{description}

\newpage
\cntr{
\begin{figure}
\cntr{\includegraphics[width=3.5in]{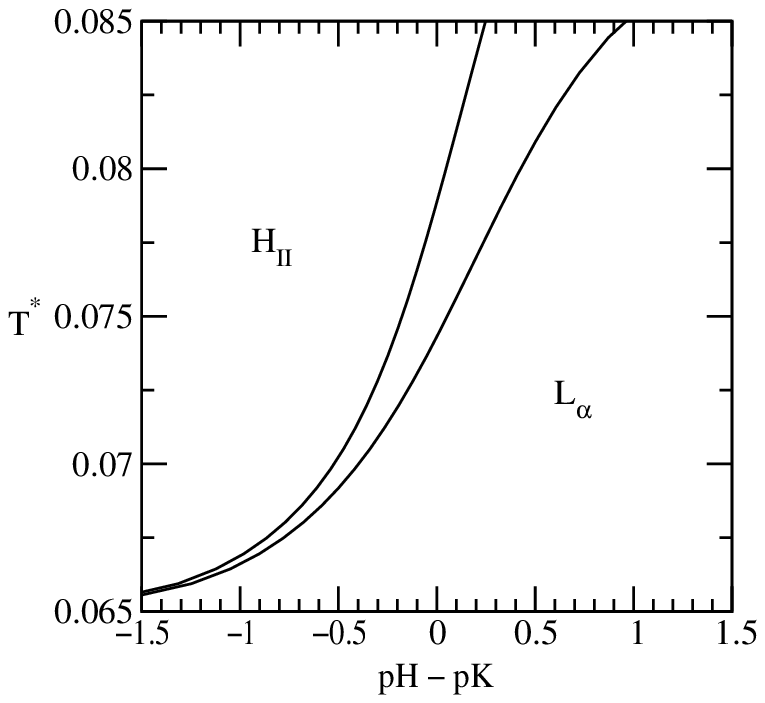}}
\end{figure}
}
\begin{center}
Figure 1: X.-J. Li and M. Schick, Biophys. J., top $\Uparrow$
\end{center}

\newpage
\cntr{
\begin{figure}
\cntr{\includegraphics[width=3.8in]{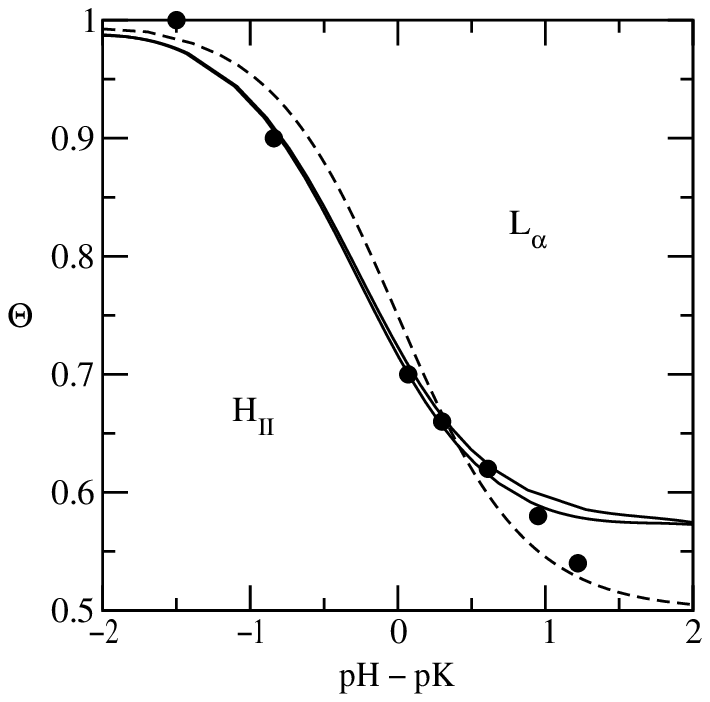}}
\end{figure}
}
\begin{center}
Figure 2: X.-J. Li and M. Schick, Biophys. J., top $\Uparrow$
\end{center}

\newpage
\cntr{
\begin{figure}
\cntr{\includegraphics[width=3.8in]{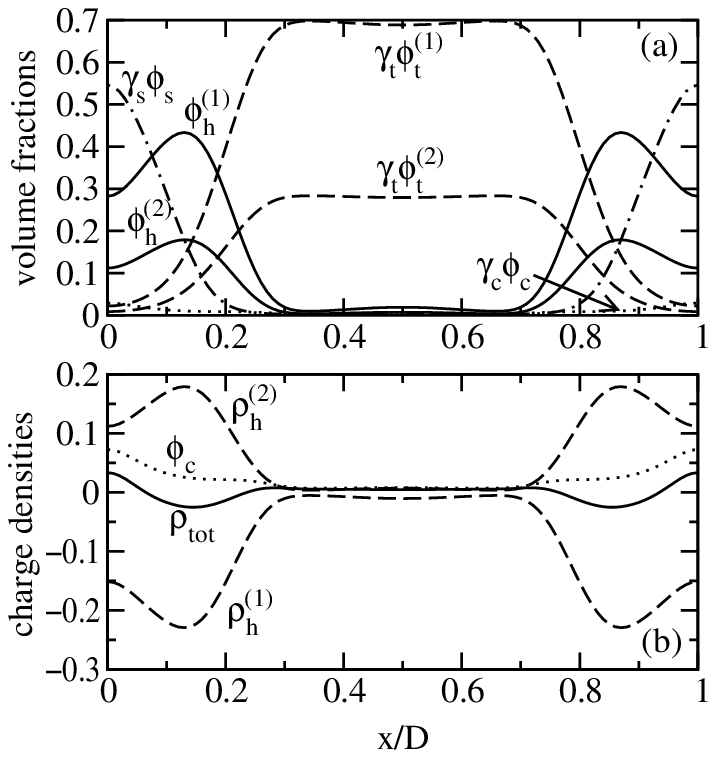}}
\end{figure}
}
\begin{center}
Figure 3: X.-J. Li and M. Schick, Biophys. J., top $\Uparrow$
\end{center}

\newpage
\cntr{
\begin{figure}
\cntr{\includegraphics[width=3.8in]{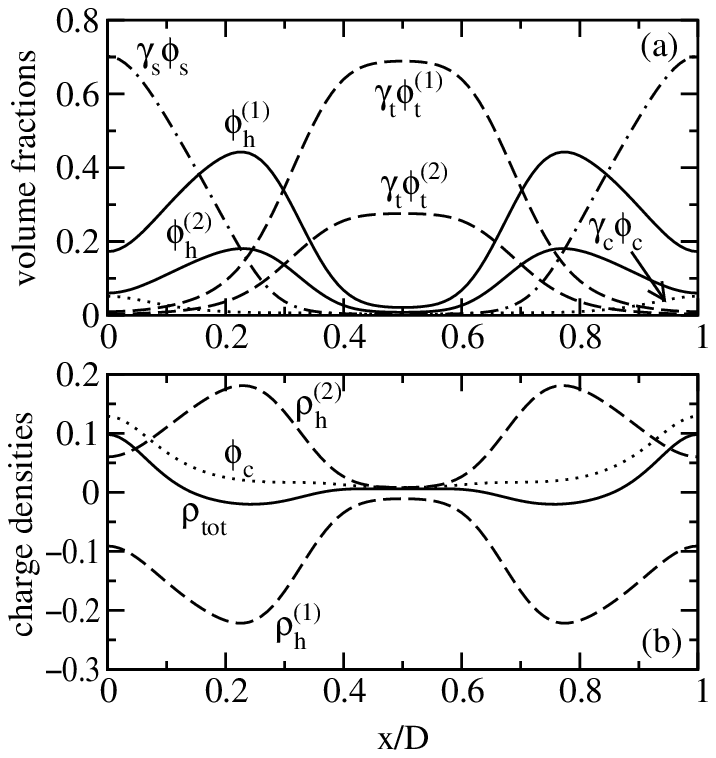}}
\end{figure}
}
\begin{center}
Figure 4: X.-J. Li and M. Schick, Biophys. J., top $\Uparrow$
\end{center}

\newpage
\cntr{
\begin{figure}
\cntr{\includegraphics[width=3.8in]{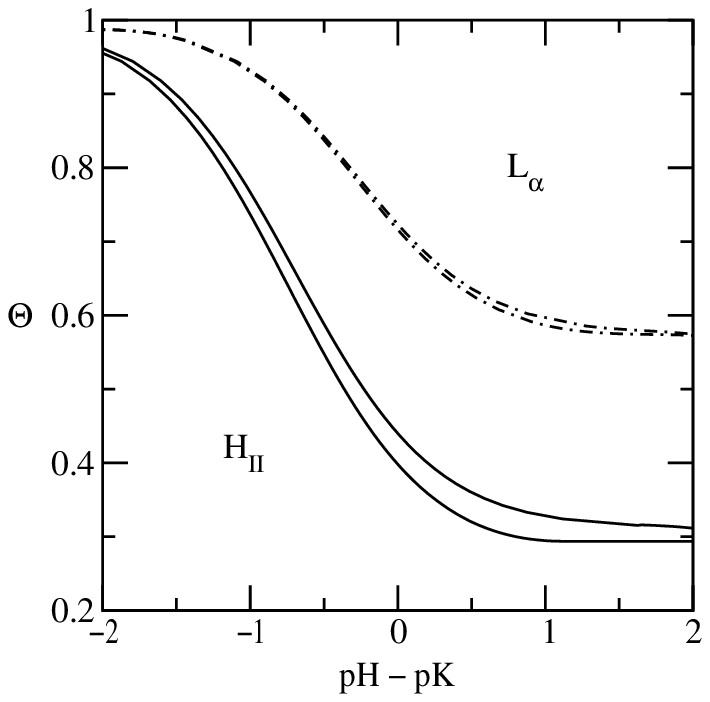}}
\end{figure}
}
\begin{center}
Figure 5: X.-J. Li and M. Schick, Biophys. J., top $\Uparrow$
\end{center}

\end{document}